# AN EFFICIENT KEY AGREEMENT SCHEME FOR WIRELESSSENSOR NETWORKS USING THIRD PARTIES

Saleh Almowuena

School of Computing Science, Simon Fraser University, Canada
salmowue@sfu.ca

**ABSTRACT**

*This paper contributes to the challenging field of security for wireless sensor networks by introducing a keyagreement scheme in which sensor nodes create secure radio connections with their neighbours depending on the aidof third parties. These third parties are responsible only for the pair-wise key establishment among sensor nodes,so they do not observe the physical phenomenon nor route data packets to other nodes. The proposed methodis explained here with respect to four important issues: how secret shares are distributed, how local neighboursare discovered, how legitimate third parties are verified, and how secure channels are established. Moreover, theperformance of the scheme is analyzed with regards to five metrics: local connectivity, resistance to node capture,memory usage, communication overhead, and computational burden.Our scheme not only secures the transmissionchannels of nodes but also guarantees high local connectivity of the sensor network, low usage of memory resources,perfect network resilience against node capture, and complete prevention against impersonation attacks. As it isdemonstrated in this paper, using a number of third parties equals to 10% of the total number of sensor nodes inthe area of interest, the proposed method can achieve at least 99.42% local connectivity with a very low usage ofavailable storage resources (less than 385 bits on each sensor node).*

**KEYWORDS**

*Key Agreement, Wireless Sensor Network, Third Party, Trust Establishment, Secure Channel.*

## 1. INTRODUCTION

Wireless sensor networks are employed in a wide range of applications including disaster relief operations, forest-fire detection, battlefield surveillance, pollution measurement, and healthcare applications. Because of the characteristics of these applications as well as the broadcast nature of the radio transmissions, a wireless sensor network is more vulnerable to security threats than traditional wireless networks. In order to protect the sensor network from outside attacks, it is necessary to implement a cryptographic mechanism that can achieve three major security objectives: confidentiality, integrity and authentication. Even though the topic of cryptography has been well studied for traditional networks, many conventional cryptographic approaches cannot easily be applied to sensor networks [1]. To illustrate, public key-based schemes and even some symmetric key methods are complex with regards to computations, memory, communication, and packet size requirements. On the other hand, sensor networks suffer from severe constraints on their available resources as a result of the necessity to increase the lifetime of the complete network, minimize the physical size of the sensor nodes, and reduce the cost of sensor nodes [2]. Consequently, it is important to propose cryptographic solutions designed specifically for wireless sensor networks.

A fundamental element in an effective cryptographic system is how sensor nodes are equipped with the cryptographic keys needed to create secure radio connections with their local neighbours. This paper contributes to the challenging field of key establishment by introducing a key agreement scheme whose memory, processing, and communication requirements are low. This method utilizes the concept of third parties to reduce the cryptographic burden of public-key based schemes and the key management overhead of symmetric key approaches. Third parties are simply additional nodes deployed in the field of interest to assist sensor nodes in the key establishment phase. Hence, they do not perform any other operations such as sensing or packet routing. Our key agreement scheme has many advantages over existing approaches. For instance, a sensor node in this scheme needs to make just a few local contacts to establish a secureradio connection with its neighbours with very high probability. In addition, the majority of sensor nodes employs just a few simple hash operations and stores only a small number of secret keys in their memory space. The proposed method also employs an authentication mechanism to prevent impersonation attacks.

To the best of our knowledge, the idea of exploiting third parties in wireless sensor networks has been discussed only in [3] where Dong and Liu used auxiliary sensors for pair-wise key establishment. However, the approach in [3] suffers from two important disadvantages: it is not scalable to redistribute additional nodes after the deployment stage, and it requires a massive amount of memory in the assisting nodes. Even though Dong and Liu state that assisting nodes can utilize all their memory to store the hash images of sensor nodes, this assumption may not be sufficient for their method to be feasible for sensor networks. To illustrate, *Telos mote* needs its entire 1 MB of flash memory to store the hash images for only 65536 sensor nodes, assuming the hash function provides a 16 bytes long value. An upper limit for the number of nodes makes the network non-scalable and so this scheme is impractical for large networks. To overcome this limitation, we propose in this work a scalable and efficient key agreement method in which both third parties and sensor nodes use a small number of memory units in the key establishment step.
The rest of the paper is organized as follows. Section 2 provides a brief survey of the current key agreement techniques in the literature. Then we start our discussion in Section 3 by presenting the trust assumptions considered in both analysis and evaluation. After that, the proposed key establishment method is described with respect to three points: how secret shares are distributed, how local neighbours are discovered, and how secure channels are created. In Section 4, we will analyze the scheme's performance using five metrics: local connectivity, resilience against node capture, memory usage, communication overhead, and computational overhead. Next, our method is concisely compared with a number of current key agreement algorithms introduced specifically for wireless sensor networks. To conclude this paper, section 5 provides some promising research directions for the future.

## 2. RELATED WORK

A simple technique to enable sensor nodes establishing secure communication channels with their neighbours is to preload a single shared key into the memory space of each node in the network. Having the same secret key provides an efficient key agreement scheme with regards to both power consumption and memory usage because sensor nodes utilize only a single unit of their available storage resources and avoid performing any data interaction. However, this method offers weak resilience against node capture since compromising any sensor node will reveal its key material and then cause a major breach in the security of the entire network. To enhance the resilience against node capture, a pair-wise key can be assigned to every pair of nodes in the network. In this case, trusted base stations should randomly generate$N(N-1)/2$ secret keys and then supply each sensor node with $N-1$ of these keys. A sensor network may have thousands to millions of nodes in the field of interest, so it is not practical to implement such approach because of the limitation in memory resources.

Under the assumption that base stations are considered trustworthy entities in wireless sensor networks, it is easy to apply the concept of Key Distribution Centre (KDC) in which base stations are responsible for assisting sensor nodes in the process of key establishment. To do so, each sensor node in the network will share a unique symmetric key with a base station. Once neighbours in its transmission range are identified, the sensor node forwards a request to the base station indicating a list of sensor nodes with which it intends to initiate secure radio connections. Responding to this request, the base station generates a set of pair-wise keys and then transmits them back to the sender. The main drawback here is the possibility of a single point of failure as well as the high number of packet transmissions.

Recently, researchers have focused on the idea of key pre-distribution in which a large pool of symmetric keys is generated before deploying sensor nodes in the target terrain [4], [5], [6], [7]. Each sensor node is equipped with a ring of secret keys chosen randomly from a key-pool. Two sensor nodes will be capable of creating a secure radio connection if both share a common key. Usually, the key-rings in pre-distribution approaches are designed in such a way that nodes can succeed with a pre-determined probability in finding shared keys with their local neighbours. Regarding the performance of the methods in this category, it has been shown that the number of symmetric keys in both the key-ring and the key-pool significantly affect the network connectivity, the resilience against node capture, and the memory usage. For example, increasing the number of secret keys in the general key-pool will enhance the security of the entire network, but it may negatively impact the local connectivity of sensor nodes. To balance this trade-off between security and connectivity, efforts have been made to optimize traditional public key-based algorithms, such as elliptic curve cryptography, in order to make them suitable for wireless sensor networks [8], [9], [10], [11].

Our method reduces the cryptographic burden of public-key based schemes and the key management overhead of random key pre-distribution approaches. To achieve this objective, the proposed scheme replaces high cost public-key operations at the sensor nodes with a few simple hash operations, that is, nodes use one-way functions in which the input is data of arbitrary length and the output is a unique value of a specific size. Furthermore, additional nodes called third parties are deployed in the network. These assisting nodes do not perform sensing, routing or packet forwarding; they are only responsible for pair-wise key establishment between sensor nodes.

## 3. An Efficient Key Agreement Scheme Using Third Parties

### 3.1. Trust Assumptions

Throughout this paper, several assumptions are considered in both analysis and discussion. In this subsection, we will highlight the major assumptions, on the basis that other complementary assumptions will be mentioned in their relevant context. First, both physical and data-link layers are vulnerable to direct attacks. Hence, wireless channels are insecure and susceptible to packet eavesdropping, data injection, message modification, and replay attacks. Second, sensor nodes are not supported with tamper-resistant modules since such components are either simple but insecure or robust but costly. As a consequence, an adversary can compromise a sensor node and extract its key materials, observation data, and software. Third, base stations are trustworthy devices, whereas sensor nodes are not guaranteed to behave reliably. Fourth, an attacker is capable of employing any of the adversary models. As an example, he can deploy malicious nodes with resources identical to those found on the sensor nodes or launch attacks taking advantage of powerful laptops. Furthermore, an attacker can compromise a number of regular nodes and manipulate them to start a collusion attack against the network. The last assumption is that an adversary targets sensor nodes randomly without prior knowledge of the keys carried on the nodes. Nevertheless, he may be aware of the security mechanisms implemented in the wireless sensor network.

## 3.2. Proposed Algorithm

Like random key pre-distribution schemes, the method introduced here consists of three main steps: distributing secret shares, discovering local neighbours, and establishing secure channels. Assume $t$ third party nodes in addition to $n$ sensor nodes are deployed uniformly in the field of interest. Prior to deployment of the sensor network, a trusted base station generates a random encryption key $S$ as well as a random authentication key $A$. Then the trusted base station will store both keys as private information into the memory units of all $t$ third parties. Furthermore, every sensor node in the network is equipped with unique encryption and authentication keys. To do so, the trusted base station computes for each node $i$ two values:

$$S_i = Hash(S, ID_i) \quad and \quad A_i = Hash(A, ID_i) \tag{1}$$

and preloads these two keys into the node memory unit.

After scattering sensor nodes in the target region, the nodes perform a discovery process in order to find their local neighbours as well as their closest third party. Finding local neighbours is a simple operation since sensor nodes periodically broadcast HELLO packets or what is called beaconing to advertise their existence in a neighbourhood. Therefore, nodes can easily depend on these packets to identify adjacent nodes. To protect the network from HELLO flood attacks [1], a pair of neighbouring nodes performs a two-way handshake to ensure that the other node is located within its radio range. On the other hand, it is not sufficient to allow third parties to advertise their existence in a neighbourhood relying on HELLO packets alone. The reason is that malicious nodes can spoof the contents of a HELLO packet originating from a genuine third party and then forward this packet to the sensor nodes in its transmission range. Innocently, these sensor nodes can be deceived to assign this malicious node as their closest third party node, thereby disrupting the functionality of the proposed method.

To verify the legitimacy of a third party in a sensor network, we utilize an authentication mechanism called hash chain scheme. This scheme provides a simple and scalable solution without consuming substantial storage resources in the sensor nodes. In this method, sensor nodes need to store just a single key in their memory units. Also, they can easily accommodate new legitimate third parties which are redistributed in the field of interest after the deployment phase. To implement the hash chain mechanism, a trusted base station randomly generates a key $M$ and then inputs this generated key in a sequence of hash functions as shown in Figure 1. Prior to deployment, third parties are given the chain of keys $L = \{L_0, L_1, L_2, \ldots, L_{a-1}, L_a\}$, while sensor nodes are equipped with a single key $L_a$. Once the sensor nodes are spread in the target region, third parties advertise their existence in a neighbourhood through broadcasting a customized HELLO packet containing the key $L_{a-1}$. Upon receiving a HELLO packet, every sensor node in the network checks the legitimacy of the third party as follows. Sensor nodes extract the key $L_{a-1}$ from the incoming HELLO packet and then input $L_{a-1}$ to the predefined hash function. If the value resulting from the hash function $H(L_{a-1})$ matches the key $L_a$ stored in node memory, authentication is successful. In this case, each sensor node should update its stored key by replacing $L_a$ with $L_{a-1}$. Next time, the third party will unitize the key $L_{a-2}$ in its customized HELLO packets, and so on. It is important to note that the value of the parameter $a$ should be chosen carefully taking into account the strength of the security protocols needed in addition to the expected lifetime of the sensor network. For instance, the parameter $a$ used in long-term battlefield surveillance should be greater than the value utilized in applications designed to obtain statistics about the number of visitors in a short-term tourist attraction.

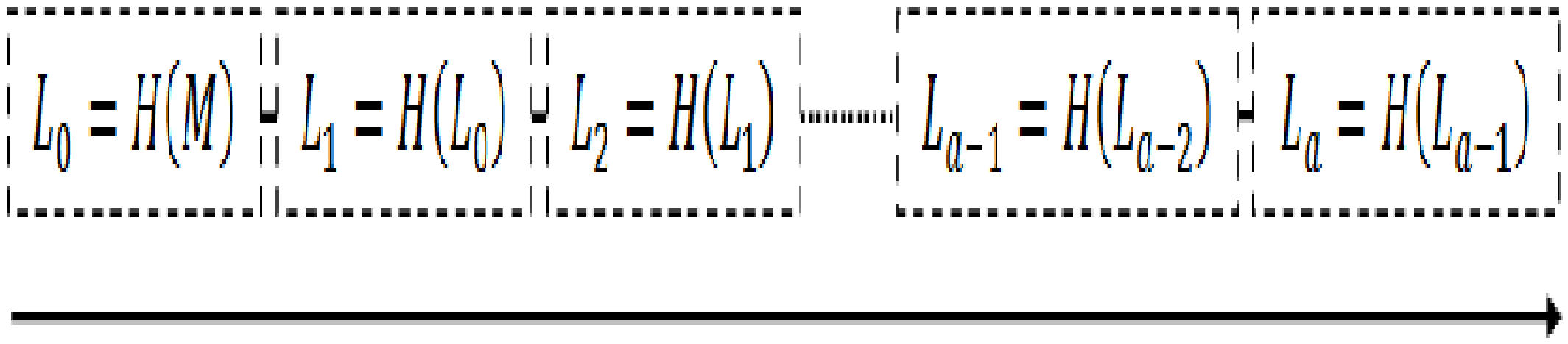

Figure 1. An example of a hash chain generated from a random key *M*

Once the discovery stage is completed, every sensor node $i$ in the network sends a request to the discovered third party in its transmission range stating its desire to establish secure channels with neighbouring nodes $\{j_1, j_2, \dots, j_d\}$. To help the third party in verifying the identity of the packet originator, node $i$ encrypts its outgoing message with the authentication secret $A_i$:

$$Node(i) \rightarrow Third\ Party: \quad E_{A_i}(Request(i, \{j_1, j_2, \dots, j_d\}))$$

Responding to such a request, the third party starts the process of key generation by computing the values $S_i$ and $A_i$using (1). For each neighbour mentioned in the received request, the third party determines an encryption key $S_{j_x}$ and then calculates a temporary secret share for use in generating a session key between both the sensor node $i$ and its neighbour $j_x$ as follows:

$$\mathrm{Secret}(i, j_x) = \mathrm{Hash}(S_i, ID_{j_x}) \oplus \mathrm{Hash}(S_{j_x}, ID_i) \qquad (2)$$

The third party sends back to the sensor node $i$ its temporary secret shares protected with the authentication key $A_i$ as shown below:

$$\mathrm{Third\ Party} \rightarrow \mathrm{Node}(i): \quad E_{A_i}(i \,||\, j_x \,||\, \mathrm{Secret}(i, j_x)) \quad \forall x \in [1, d]$$

Upon receiving these temporary secret shares, the sensor node $i$ uses the identity of its neighbouring node $j_x$ in order to determine a session key between the two nodes given by:

$$\mathrm{Session}_{j_x} = \mathrm{Secret}(i, j_x) \oplus \mathrm{Hash}(S_i, ID_{j_x}) \qquad (3)$$

Although sensor nodes $i$ and $j_x$ can rely on this session key to secure their radio connections, we recommend that sensor node $i$ generates a new random secret key $K_{i,j_x}$ and then constructs a packet containing this random key. This packet should be encrypted with the session key obtained in (3) and then forwarded to the neighbouring node $j_x$. Performing this additional step boosts the network security and prevents any negative impact from the disclosure of session keys after a third party is compromised. As soon as the neighbouring node $j_x$ receives the encrypted packet, it inputs the identity of node $i$ into the predefined hash function in order to extract the session key used to encrypt the received packet as follows:

$$Session_i = Hash(S_{j_x}, ID_i) \qquad (4)$$

Obviously, the session keys in (3) and (4) are identical. To prove this point, we can easily substitute (2) into the right hand of (3) yielding (4).At this point, the neighbouring node $j_x$ uses the session key in (4) to decrypt the received packet and then obtain the secret key $K_{i,j_x}$. Sensor nodes $i$ and $j_x$ can then employ the secret key $K_{i,j_x}$ to secure their radio channel.

Since it is possible to encounter situations in which a sensor node $i$ has no third parties in its local neighbourhood, it is necessary to include an additional step in the key establishment phase to avoid such scenarios. In this step, a neighbour of node $i$ will help him in finding a third party within a predefined number of hops and then communicate with this third party on behalf of node $i$. Similar to what was explained earlier, sensor node $i$ will send a request to a neighbour indicating its desire to establish secure channels with nodes $\{j_1, j_2, \dots, j_d\}$. This request will be encrypted with the authentication secret $A_i$ in order to protect the contents of the packet and confirm the identity of the originator. The neighbouring node will forward this request to a nearby third party in its local transmission range. Upon receiving the anticipated responses from the third party, the neighbouring node will forward these responses to node $i$. Once the temporary secret shares are received, node $i$ will continue the process of key establishment so that secure channels are created with its adjacent nodes.

## 4. EVALUATION

### 4.1. Local Connectivity

The local connectivity of a sensor network is typically represented by the probability of two neighbouring nodes being able to find a common secret key. Since sensor nodes in our proposed method depend on the assistance of a third party to generate shared keys with their neighbours, the definition of local connectivity should be different. In other words,the local connectivity of a sensor network in the proposed method is the probability that one of two neighbouring nodes discovers a third party in its neighbourhood.To compute $p_{Local}$, it is necessary to: a) determine the expected area of coverage for two adjacent nodes in the deployment region, and b) calculate the probability that at least one third party is located within this area. In the following subsections, we assume that sensor nodes in addition to third parties are distributed uniformly over a field of interest whose size is equal to $G$.

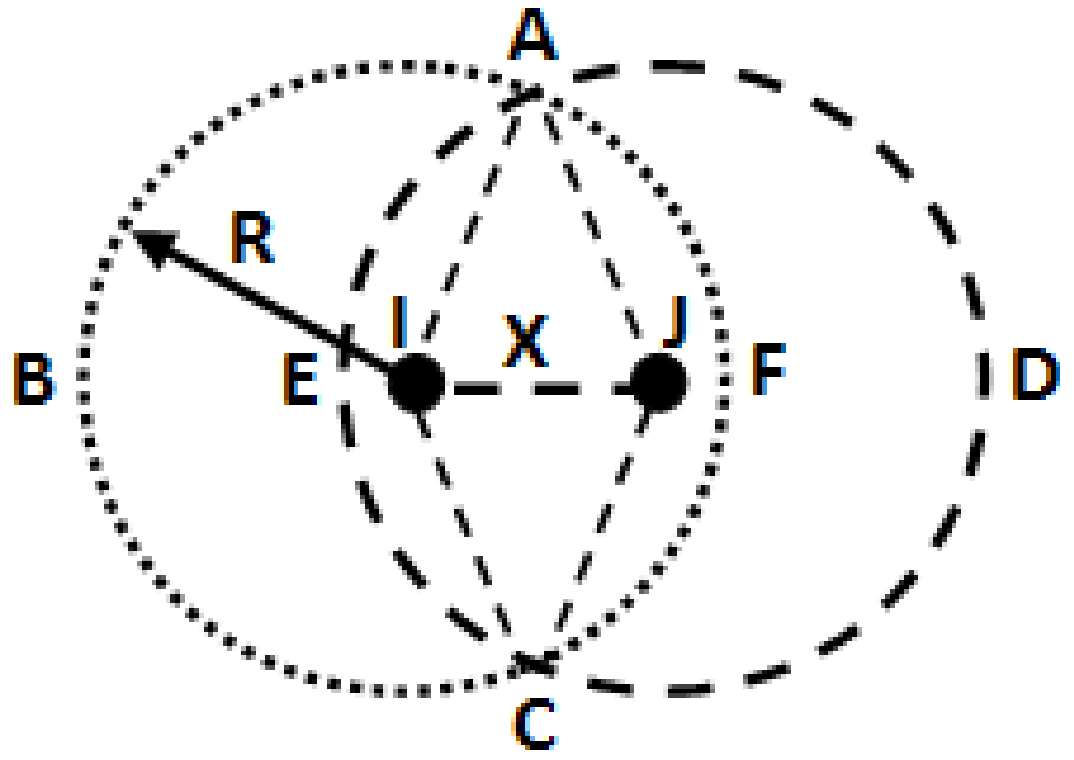

Figure 2.A pair of neighbouring nodes in a wireless sensor network

Figure 2 shows a pair of neighbouring nodes in a wireless sensor network. According to Chan et al. in [4], the expected area of coverage for the two neighbouring sensor nodes $i$ and $j$ is$\breve{E}(x) = \int_0^R \text{Area}_{ABCD}(x)\, f(x)\, dx$where $\text{Area}_{ABCD}(x)$ represents the area of both circles minus the overlapped region $AECF$, $R$ indicates the transmission radius of a sensor node, and $f(x)$ is the probability density function of the sensor node distribution in the field of interest. Because sensor nodes and third parties are distributed uniformly in the deployment area, $f(x)$ is calculated as$f(x) = \frac{d}{dx}\left(P(\text{distance} < x)\right) = \frac{d}{dx}\left(\frac{\pi x^2}{\pi R^2}\right) = \frac{2x}{R^2}$.Regarding$\text{Area}_{ABCD}(x)$, we must first compute the overlapped region AECF which is given by $Area_{AECF}(x) = 2R^2\cos^{-1}\left(\frac{x}{2R}\right) - x\sqrt{R^2 - \frac{x^2}{4}}$. Then subtracting the overlapped region from the area of both circles in Figure 2 yields $Area_{ABCD}(x) = 2\pi R^2 - 2R^2\cos^{-1}\left(\frac{x}{2R}\right) + x\sqrt{R^2 - \frac{x^2}{4}}$. On this ground, the expected area of coverage for sensor nodes $i$ and $j$ is given by:

$$\breve{E}(x) = \int_0^R Area_{ABCD}(x)\, f(x)\, dx = 1.413497\pi R^2$$

Since third parties are uniformly deployed in the wireless sensor network, the probability that a third party node is located inside the region $ABCD$ as follows:

$$p = \frac{\text{The expected region } Area_{ABCD}(x)}{\text{The size of the deployment area}} = \frac{\breve{E}(x)}{G} = \frac{1.413497\pi R^2}{G}$$

Because it is possible that more than one third party is located in a particular area, the binomial distribution can be used to derive the probability that $z$ third parties are within the region $ABCD$. This probability is given by $p(z = \breve{z}) = \binom{t}{\breve{z}} p^{\breve{z}}(1-p)^{t-\breve{z}}$. As mentioned earlier, a pair of neighbouring nodes can establish a secure radio connection when one of the two nodes has at least one third party in its transmission range. Based on this, we can conclude that the local connectivity of a sensor network in the proposed method is given by:

$$p_{Local} = p(z \geq 1) = 1 - p(0) = 1 - \left(\binom{t}{0} p^0 (1-p)^t\right) = 1 - (1-p)^t \quad (5)$$

To simplify (5), let $d$ be the average number of neighbours within the radio coverage of a sensor node. When $n \square d$ and $G \square \pi R^2$, we can assume $\frac{\pi R^2}{G} = \frac{d}{n}$. Then the local connectivity can be rewritten as follows:

$$p_{Local} = 1 - (1-p)^t = 1 - \left(1 - \frac{1.413497\pi R^2}{G}\right)^t = 1 - \left(1 - \frac{1.413497\, d}{n}\right)^t \quad (6)$$

Figure 3 shows the local connectivity of a network when the number of sensor nodes is equal to 10000 with various densities, i.e., the average number of neighbours in a node's transmission range. Because the number of third party nodes plays an important role in determining the local connectivity of a network, we plot the local connectivity with respect to the ratio of the number of third parties to the number of sensor nodes. Clearly, increasing this ratio to more than 40 percent will result in perfect connectivity, but this is not feasible from both practical and economical points of view. As a consequence, we consider lower ratios such as 10 percent which provides reasonable connectivity. As shown in Figure 3, the proposed method with a 10 percent ratio gives approximately 92.5 percent of local connectivity even in low dense networks like when $d = 20$.

Figure 3 also shows it is possible that a pair of neighbouring nodes may have no third party in their transmission range. As a result, it will be necessary to depend on an intermediate node in order to complete the key establishment process, as it was described previously. Since the communication overhead is an important concern in wireless sensor networks, we should limit the number of intermediate nodes between a sensor node and a third party to only one. In other words, if two neighbours are trying to establish a pair-wise key to secure their radio channel, one of the two nodes should discover a third party within two hops as shown in Figure 4. Note that a two hop distance does not necessary mean a radius of 2R. This value is quite optimistic in non-dense sensor networks as it would require the intermediate node to be located on the edge of a neighbouring node boundary. To compute the local connectivity through an intermediate node in a non-dense sensor network, a good approximation for the two hop range is a radius of 3R/2 rather than 2R.

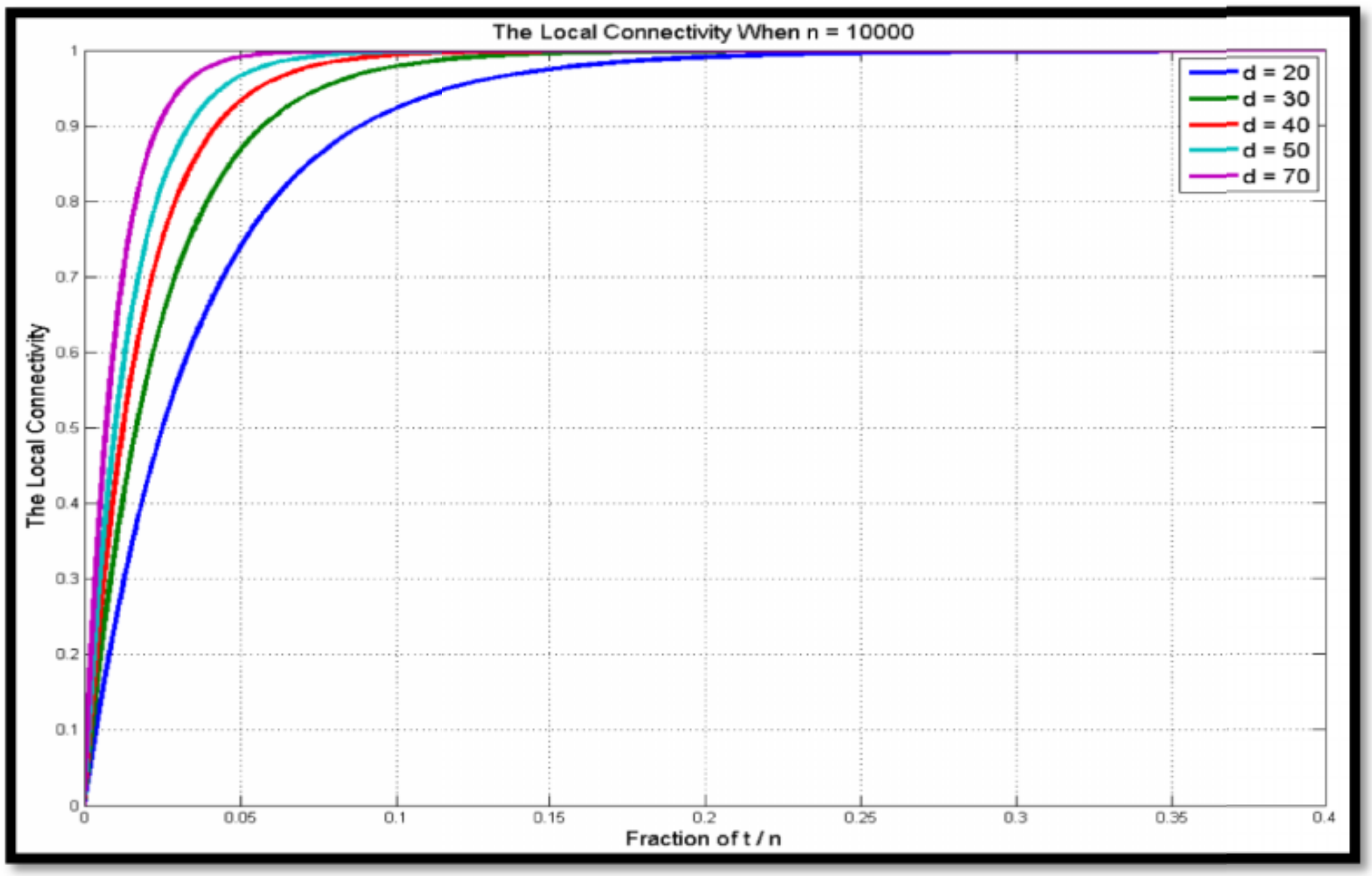

Figure 3.The local connectivity of a network applying our method

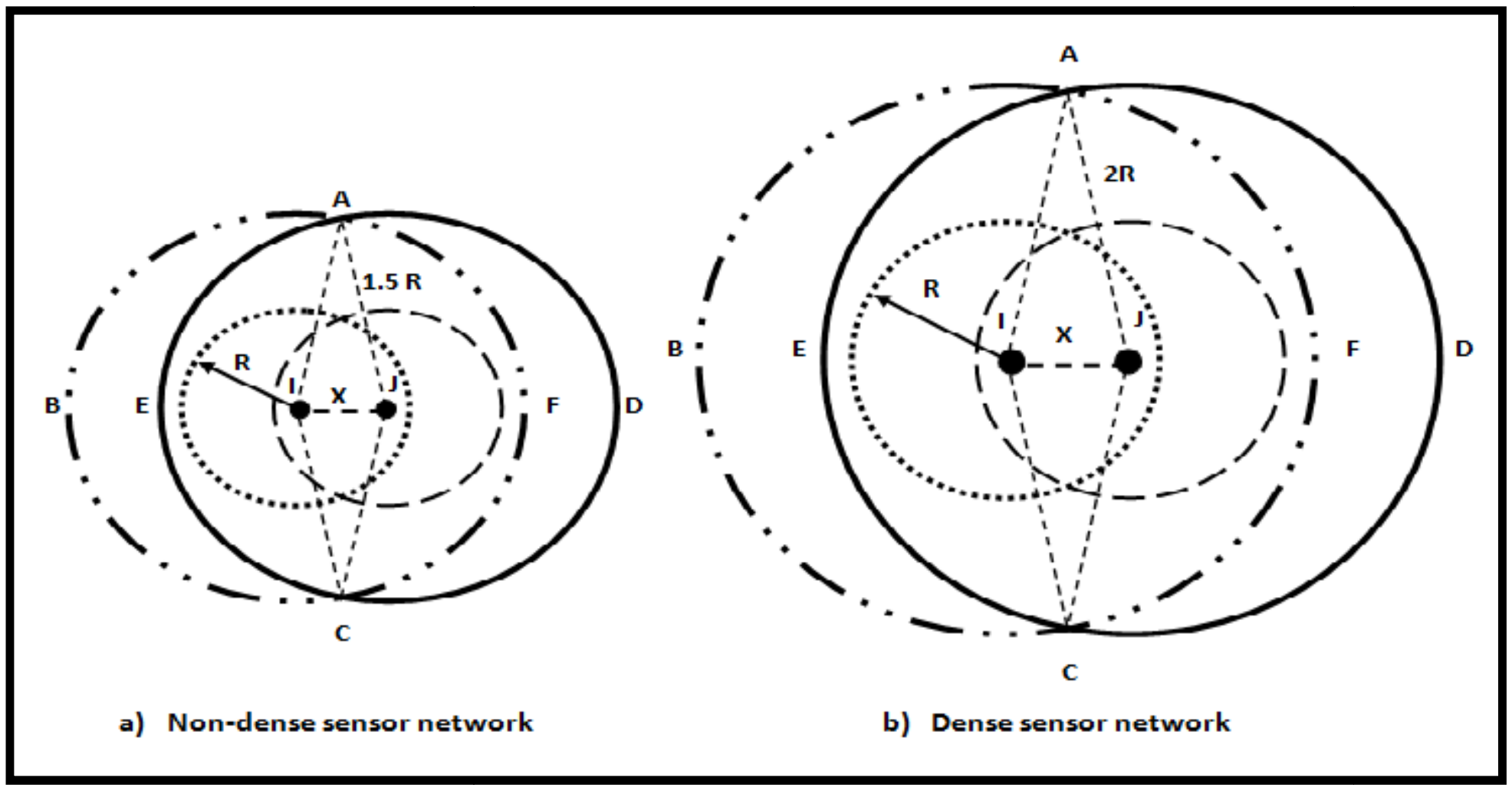

Figure 4. The scope of discovery possible for two neighbouring nodes

Allowing sensor nodes to seek assistance from an intermediate node to discover a third party can increase the local connectivity of a sensor network. To calculate $p_{Local}$ in this case, we simply need to compute the expected area of ABCD in Figure 4 and then determine the probability that at least one third party is located inside this region. In situations where sensor nodes are deployed in a non-dense network, $Area_{ABCD}$ represents the area of both circles in Figure 4.a minus their overlapped region AECF as follows:

$$Area_{ABCD}(x) = 4.5\pi R^2 - 4.5\,R^2 \cos^{-1}\left(\frac{x}{3R}\right) + x\sqrt{\frac{9}{4}R^2 - \frac{x^2}{4}}$$

Therefore, the expected area of coverage for two neighbouring nodes in a non-dense network and the probability that a third party node is located inside ABCD are $\breve{E}(x) = 2.87947\pi\, R^2$ and $p = 2.87947\pi\, R^2/G$, respectively. Similar to what was done earlier, the binomial distribution can be used in order to determine the probability that at least one third party is located within the region ABCD. Based on this, the local connectivity of the non-dense sensor network in Figure 4.a is given by:

$$p_{Local} = 1 - (1-p)^t = 1 - \left(1 - \frac{2.87947d}{n}\right)^t \qquad (7)$$

On the other hand, the region $ABCD$ in the dense sensor network in Figure 4.b is as follows:

$$Area_{ABCD}(x) = 8\pi R^2 - 8\,R^2 \cos^{-1}\left(\frac{x}{4R}\right) + x\sqrt{4R^2 - \frac{x^2}{4}}$$

Consequently, the expected area of coverage for sensor nodes i and j in Figure 4.b is $\breve{E}(x) = 4.84349\pi\, R^2$ making the local connectivity of the dense network to become:

$$p_{Local} = 1 - (1-p)^t = 1 - \left(1 - \frac{4.84349d}{n}\right)^t \qquad (8)$$

In this subsection, three different scenarios were considered in the key establishment phase: a) sensor nodes are not allowed to use any intermediate node, b) sensor nodes are deployed in non-dense networks in which they can use a maximum of one intermediate node, and c) sensor nodes are distributed in dense networks where they can utilize at most one intermediate node. When the average number of neighbours within the transmission range of a sensor node is equal to $d = 20$, Figure 5 illustrates the improvement obtained using a maximum of one intermediate node in our proposed method. For example, the network in scenario (a) needs 4.72 times the number of third parties required in scenario (c) such that a fully connected network is achieved. Compared with the non-dense networks in scenario (b), the network in scenario (a) still requires 2.59 times the number of third parties needed in scenario (b) in order to have a complete local connectivity. Changing $d$ from 20to 40, the performance will be enhanced since increasing the expected number of neighbours within a transmission range boosts the probability that a pair of neighbouring nodes can discover a third party. Figure 6 shows a comparison between the local connectivity of the three scenarios when $d = 40$.

## 4.2. Resilience against Node Capture

When an adversary captures a node in a wireless sensor network, he is able to extract the key material, observation data, and software stored on that node. Under this circumstance, it is important to employ key agreement schemes for which a compromised sensor node has a limited impact on the security of the entire network. In other words, the secret keys of non-compromised nodes should not be revealed when a sensor node is captured. Usually, a metric called resilience against node capture is utilized in order to measure the resistance of a key agreement scheme to a node being compromised. This metric is calculated as the ratio of the number of compromised nodes to the percentage of insecure radio connections in the network. High resilience against node capture indicates that a compromised sensor node has a low impact on the secrecy of transmission channels belonging to other nodes.

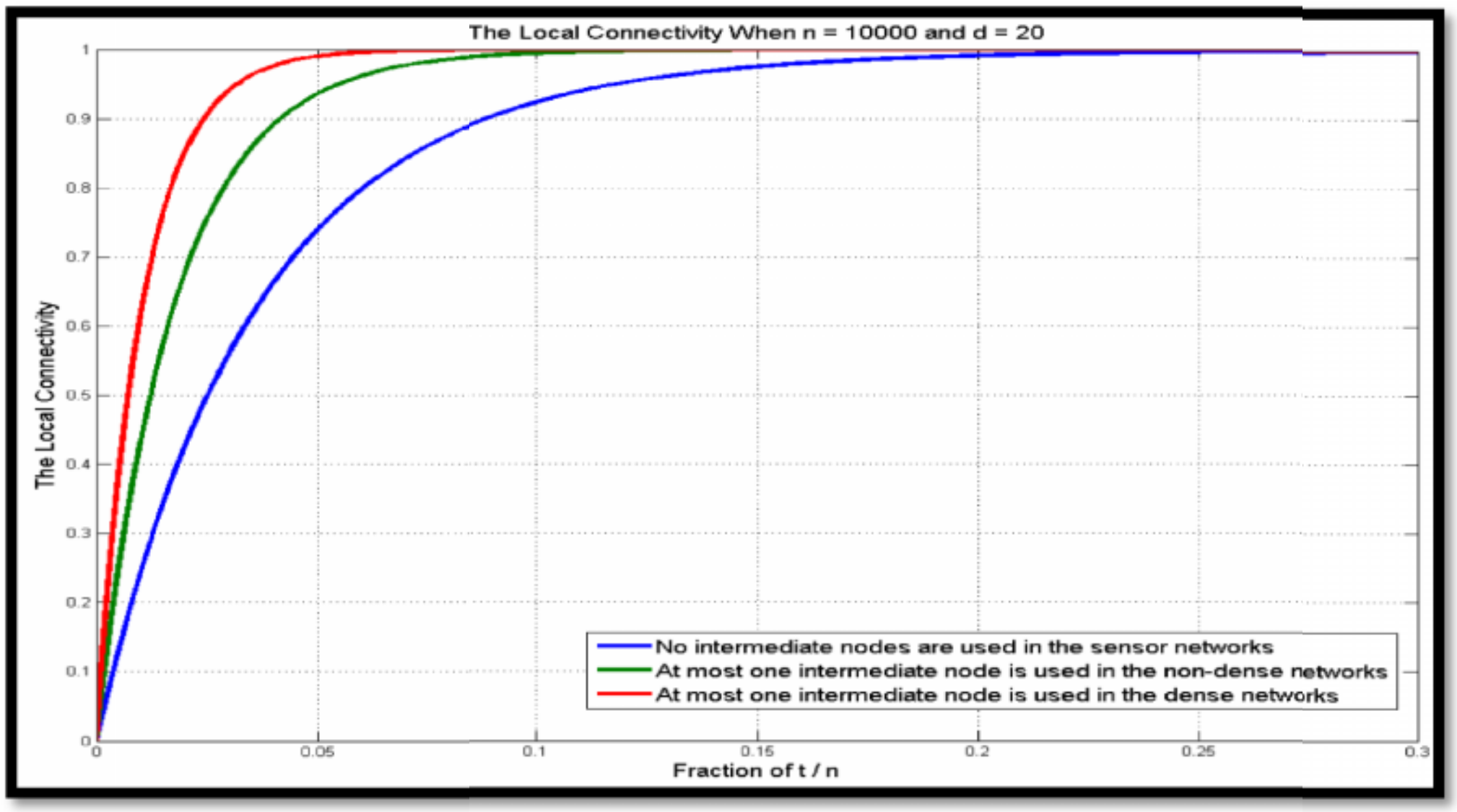

Figure 5.The local connectivity for three scenarios when d = 20

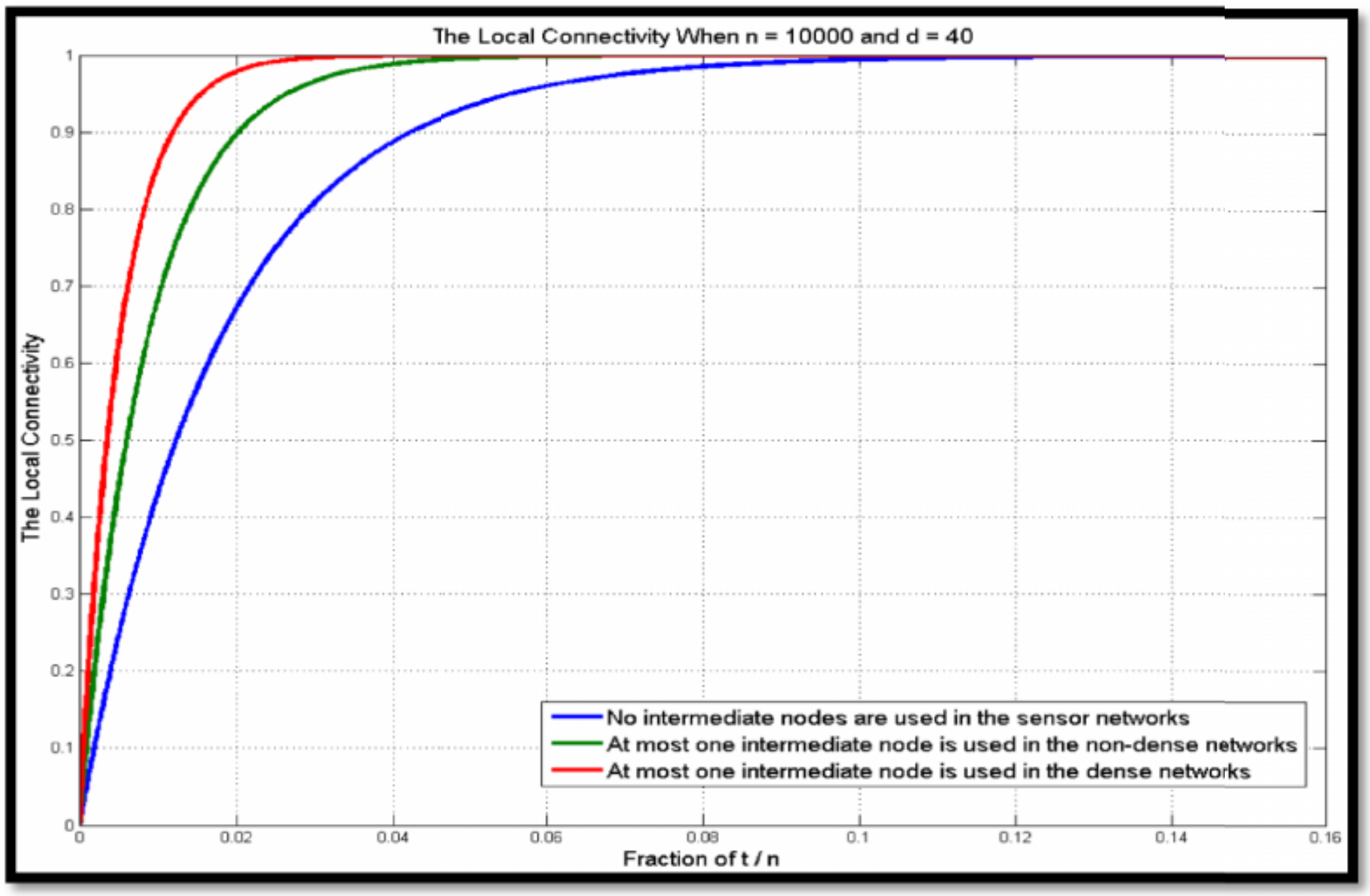

Figure 6.The local connectivity for three scenarios when d = 40

Unlike the majority of random pre-distribution schemes, compromising a sensor node in our proposed method does not have any effect on the secrecy of transmission channels belonging to other nodes in the network. To clarify, sensor nodes in our scheme store only their private information, which is independent from the private data in other nodes. Consequently, the exclusive use of private keys in the network confines the negative impact of a compromised sensor node to the node itself. On the other hand, compromising a single third party may lower, in rare situations, the resistance of the proposed scheme to a node being compromised. To illustrate, every pair of neighbouring nodes depends on their closest third party to establish a secure radio connection. The third party generates a session key which is used by the pair of sensor nodes to exchange a random secret key. Even though it would be easier for the two neighbouring nodes to rely on the session key generated by their closest third party for communications, the proposed scheme makes both nodes secure their transmission channel with a random secret key of their own. This measure prevents the negative impact caused by the disclosure of session keys after a third party is compromised. Therefore, the secrecy of non-compromised radio connection is preserved. However, it is still possible in some situations that a third party becomes compromised while pairs of sensor nodes are still exchanging their random secret keys. In this scenario, the radio channels of these nodes may be compromised leading to a slight decrease in the network resilience against node capture.

A reasonable solution to reduce the negative impact of a compromised third party would be deleting the private information stored in the memory units of third parties as soon as the key agreement process is completed. Normally, the key establishment phase is performed in the beginning of a network lifetime, and it is often accomplished within a period of time $T_k$. Wireless sensor networks are typically static configurations in which nodes do not change their locations after deployment. On this basis, a sensor node is expected to not seek any further assistance from third parties after the period $T_k$ unless new nodes are redistributed in its neighbourhood. Under these conditions, deleting the private information stored on third parties after $T_k$ and then putting these third parties in sleep mode will reduce the risk of capturing a third party without affecting the local connectivity of a sensor network. We can also assume that attackers require time larger than $T_k$in order to capture a third party. With this assumption, our proposed method provides perfect resilience against node capture. When the network redeploys additional nodes in the field of interest, third parties can be awakened to receive from the trusted base station an encrypted packet containing their private information. Completing the key agreement process, these third parties will delete their secret keys within the period $T_k$and then go back to sleep mode.

## 4.3. Memory Usage

Memory usage is a metric defined as the amount of memory that a node requires in order to establish secure transmission channels with its neighbours. A significant characteristic of the proposed method is its small memory usage in sensor nodes as well as third parties. To illustrate, every sensor node in the network starts the key agreement process by discovering its closest third party. For this step, sensor nodes need to store just a single key in their memory space so that the hash chain mechanisms can be performed. After discovering neighbouring third parties, nodes should continue the key establishment phase so that their radio connections are secured. To accomplish this, each sensor node is preloaded with unique encryption and authentication keys. Hence, sensor nodes in the proposed scheme must store only three keys.

Similar to sensor nodes, a third party needs to store three secret keys: an encryption key $S$ to assist a pair of neighbours in generating session keys, an authentication key $A$ to identify the originators of packets sent to or from sensor nodes, and a random value $M$ to help sensor nodes in verifying the legitimacy of a third party in their region. In addition to these three keys, a third party should also be equipped with an exclusive pair-wise key used to initiate secure routes to trusted base stations. As a consequence, third parties in the proposed method require four secret keys in order to conduct the key agreement process. Assuming cryptographic keys are $128$ bits in length [12], the memory usage of sensor nodes and third parties are $384$ and $512$ bits, respectively.

## 4.4. Communication and Computational Overhead

Another important metric to evaluate the performance of key agreement schemes for wireless sensor networks is the complexity of communication and processing operations. In this regard, our proposed method has many advantages over existing public key and random pre-distribution techniques. For example, it replaces high cost public-key operations at the sensor nodes with symmetric encryption and a few hash operations. Furthermore, a sensor node needs to conduct just a few local contacts in order to establish pair-wise keys with its neighbours. In this subsection, both communication and computational overhead of the proposed scheme is analyzed. To secure the radio connections of a network, sensor nodes and third parties collaborate with each other to generate an exclusive pair-wise key for each pair of neighbouring nodes. Due to this cooperation, some of the computational operations required in the key establishment process can be performed on sensor nodes, while other operations can be executed on third parties. Here, it is assumed that a sensor node is responsible for generating $d/2$ secret keys on average, where $d$ is the expected number of nodes in a neighbourhood. The reason behinds choosing this value is that a pair of neighbouring nodes can establish a secure radio channel if one of the two nodes obtains assistance from a third party in its transmission range. Thus, a sensor node on average is involved in generating half of its secret keys and receives the remaining half from its neighbours. Also, it is not taken into consideration the communication overhead in discovering local neighbours. Since this process is an essential step in all routing protocols, it would be reasonable to assume that sensor nodes know the identities of neighbours prior to the beginning of the key agreement phase. To secure its radio connections, a sensor node needs to encrypt and send $d$ packets, perform the hash function $(d + 1)$ times, receive and decrypt $d$ packets, and use the random number generator $d/2$ times. On the other hand, to help a node in generating session keys with its neighbours, a third party receives and decrypts $d/2$ packets, performs the hash function $(d + d/2 + 3)$ times, encrypts $d/2$ packets, and sends $d/2$ packets back to the sensor node. To indicate the computational and communication overhead for these operations, Table 1 shows the amount of energy consumed by each operation when a sensor node is equipped with a 4MHz 8-bit Atmel ATmega128L microcontroller and a 915MHz low-power radio transceiver [13], [14].

Table 1.The energy consumed by computational and communication operations.

| Operation | Energy Consumed |
|---|---|
| Encryption using AES-128 | $1.62$ μJ/Byte |
| Decryption using AES-128 | $2.49$ μJ/Byte |
| Hashing using SHA-1 | $5.90$ μJ/Byte |
| Generating a cryptographic key | $11.4$ μJ/Byte |
| Receiving a packet | $28.6$ μJ/Byte |
| Transmitting a packet | $59.2$ μJ/Byte |

## 4.5. Analysis and Discussion

Different from the proposed method, the key establishment schemes introduced in [4], [5], [6], and [7] suffer from a poor trade-off between connectivity, security, and memory usage. For example, it has been shown that the number of symmetric keys stored on a sensor node affects the performance of the *basic scheme* in [6]. To illustrate, increasing the number of secret keys enhances the local connectivity of the sensor network, but negatively impacts the resilience against node capture and obviously increases the memory usage in sensor nodes. For instance, the key agreement process in [6] consumes 320 Bytes of node memory and results in a low local connectivity of 3.93% if every sensor node is equipped with 20 keys. In such a situation, capturing 11.50% of the sensor nodes will compromise 90% of the secure radio links in the network. When the number of symmetric keys stored on a sensor node is increased to 100 keys, this process will consume 1600 Bytes of memory but enhance the local connectivity of the network by a factor of around 12to 47.54%. However, the resilience against node capture in this case is much lower allowing an adversary to compromise 100% of the secure radio connections once 9.86% of the sensor nodes are captured.

According to Liu and Ning in [7], the performance of the multiple-space matrix pre-distribution scheme [5] and the multiple-bivariate polynomial pre-distribution scheme [7] are equivalent. Based on this ground, it is sufficient to focus the discussion here on the method in [5]. The local connectivity of the multiple-space scheme as well as its resistance to node capture are considerably influenced by two parameters: $\omega$ which denotes the number of private matrices in the network and $\tau$ which represents the number of unique vectors chosen randomly for each node. To decrease the local connectivity of a network but improve its resilience against node capture, trusted base stations should increase the number of private matrices in the network and decrease the number of unique vectors chosen for each sensor node. Similar to the*basic scheme*, the number of secret keys stored on a sensor node also plays a significant role in determining the performance of the multiple-space method. This number influences the memory usage and the resistance of the network to compromised nodes, but it does not affect the local connectivity. For example, the key establishment in [5] ensures a high local connectivity of 99.6% and utilizes 1600 Bytes of storage resources if 10 private matrices are available and 5 unique vectors are selected for each node. Yet, the security resilience in this situation is very low since capturing 0.80% of the sensor nodes will break the security of all radio connections in the network. On the other hand, increasing the number of private matrices to 90 improves the network resistance to compromised nodes by 13, yet its local connectivity is reduced approximately by a factor of 6 leading to a local connectivity of only 16.89%.

The random pair-wise key scheme [4] is designed to ensure perfect resilience against node capture. To accomplish this objective, every cryptographic key in the sensor network is exclusively used to secure a single transmission channel. Consequently, compromising a sensor node reveals only its own key material without affecting the secrecy of the radio connections belonging to other nodes. However, this method suffers from a great trade-off between memory usage and local connectivity. To increase the local connectivity of a sensor network, nodes need to consume most of their available storage resources.

Dong and Liu introduced a key establishment scheme [3] which is similar to our proposed method from the perspective that both techniques employ the concept of third parties in wireless sensor networks. Comparing the two schemes, we indicate that the local connectivity of both methods should be identical given that third parties as well as sensor nodes are uniformly deployed in the field of interest. However, they differ because the expressions used to determine the local connectivity in [3] are only estimates. Regarding the memory usage, every sensor node using the key agreement scheme in [3] needs to store a single secret key in its storage space. On the other hand, third parties in [3] need a massive amount of memory for key establishment in the network. Not equipping third parties with large memory units eventually makes the scheme in [3] impractical for large networks. The Dong and Liu's scheme is quite resistance to node capture for two reasons: only public information is stored on third parties and private keys are known only to the corresponding sensor nodes. Still, capturing a third party may result in compromising the radio connections of nodes whose session keys were generated by that third party. Once a third party is captured, the adversary can obtain its random number generator and then retrieve the session keys previously generated.

In short, our proposed method has important advantages compared with the schemes in [3], [4], [5], [6], and [7]. For instance, the proposed scheme not only secures the transmission channels of nodes but also guarantees high local connectivity of the sensor network, low usage of memory resources, and perfect network resilience against node capture. Using a number of third parties equal to 10% of the number of sensor nodes in the field of interest, the proposed scheme achieves 99.42% local connectivity when the expected number of nodes in a neighbourhood is equal to 40, as shown in Figure 3. In this case, sensor nodes need to store only 48 Bytes in their memory. Moreover, capturing a sensor node does not compromise the radio connections of non-compromised nodes, thus providing perfect resilience against node capture.

## 5. CONCLUSION

This paper focuses on the issue of key establishment for wireless sensor networks in which nodes and third parties are uniformly deployed in the field of interest. To secure the radio channel of two neighbouring nodes, one of the two communicates with a third party in its transmission coverage area asking for a session key. Furthermore, this sensor node randomly generates an exclusive secret key and constructs a packet containing this key. The packet is next encrypted by the session key received from the third party and then forwarded to the neighbouring node. It was demonstrated in this paper that our scheme does not only secure the radio connections of sensor nodes but also provides high local connectivity for the network, low usage of memory resources, and perfect network resilience against node capture. Using a number of third parties equal to five percent of the total number of sensor nodes in the area of interest, our method achieves 93.28% local connectivity if the number of sensor nodes in a neighbourhood is equal to 40 on average. In this case, a sensor node needs to store only 384 bits in its memory. Also, capturing a sensor node or a third party has no negative impact on the radio connections of other non-compromised nodes if the private keys belonging to third parties are deleted within $T_k$ seconds from the beginning of the key agreement phase.

Several possible future research directions can be derived from the work presented here. For instance, we focused on proposing efficient and scalable key agreement scheme which allow a pair of neighbouring nodes to share a unique pair-wise key. Nevertheless, we have not mentioned explicitly any particular procedure to detect and revoke the identities of compromised sensor nodes or third parties in the field of interest. Consequently, it is desirable to provide a simple mechanism that helps in detecting misbehaving nodes and third parties. These captured entities can then be isolated from the sensor network by denying them the ability to exchange messages with authorized units. Clearly, the matter of discovering a compromised entity can be considered an anomaly intrusion detection problem. Since conventional intrusion detection systems are usually complex and consume a significant amount of node resources, it would be important to introduce a simple detection and revocation model suitable for the characteristics of wireless sensor networks.

## ACKNOWLEDGEMENT

The author would like to thank ProfessorAaron Gulliver for his insights and valuable feedback that contributed significantly to this work. Sincere appreciation is also expressed to King Saud University in Riyadh, the Ministry of Higher Education in Saudi Arabia, and the Saudi Arabian Cultural Bureau in Canada for their financial support.

## REFERENCES

[1] C. Karlof and D. Wagner, (2003)“Secure routing in wireless sensor networks: attacks and countermeasures”,Ad Hoc Networks, Vol. 1, No. 23, pp. 293–315.

[2] D. Puccinelli and M. Haenggi, (2005)“Wireless sensor networks: Applications and challenges of ubiquitous sensing”,IEEE Circuits and Systems Magazine, Vol. 5, No. 3, pp. 19–29.

[3] Q. Dong and D. Liu, (2007)“Using auxiliary sensors for pairwise key establishment in WSN”, inNETWORKING 07: Ad Hoc and SensorNetworks, Wireless Networks, and Next Generation Internet, pp. 251–262, Berlin, Germany.

[4] H. Chan, A. Perrig, and D. Song, (2003)“Random key predistribution schemes for sensor networks”, in IEEE Symposium on Research in Security and Privacy, pp. 197–213, Oakland, CA.

[5] W. Du, J. Deng, Y. Han, and P. Varshney, (2003)“A pairwise key pre-distribution scheme for wireless sensor networks”, in the 10th ACM conference on Computer and Communications Security, pp. 42–51, Washington, DC.

[6] L. Eschenauer and V. D. Gligor, (2002) “A key-management scheme for distributed sensor networks”, in the 9th ACM conference on Computer and Communications Security, pp. 41–47, Washington, DC.

[7] D. Liu and P. Ning, (2003)“Establishing pairwise keys in distributed sensor networks”, in the 10th ACM conference on Computer and Communications Security, pp. 52–61, Washington, DC.

[8] E. Blaβ and M. Zitterbart, (2005) “Towards acceptable public-key encryption in sensor networks”, in the ACM Second International Workshop on Ubiquitous Computing (ACM SIGMIS), pp. 1–6.

[9] G. Gaubatz, J. Kaps, and B. Sunar, (2004)“Public keys cryptography in sensor networks – revisited”, in the 1st European Workshop on Security in Ad-Hoc and Sensor Networks (ESAS), pp. 2–18, Heidelberg, Germany.

[10] Q. Huang, J. Cukier, H. Kobayashi, B. Liu, and J. Zhang, (2003)“Fast authenticated key establishment protocols for self-organizing sensor networks”, in the 2nd ACM International Conference on Wireless Sensor Networks and Applications, pp. 141–150, San Diego, CA.

[11] D. Malan, M. Welsh, and M. D. Smith, (2004)“A public-key infrastructure for key distribution in tinyos based on elliptic curve cryptography”, inthe First Annual IEEE Conference on Sensor and Ad Hoc Communications and Networks, pp. 71–80, Santa Clara, CA.

[12] D. Carman, P. Kruus, and B. Matt,“Constraints and approaches for distributed sensor network security”, Technical Report 00010, NAILabs.

[13] A. Francillon and C. Castelluccia, (2007) “Tinyrng: A cryptographic random number generator for wireless sensors network nodes”, in the5th International Symposium on Modeling and Optimization in Mobile, Ad Hoc and Wireless Networks, pp. 1–7, Limassol, Cyprus.

[14] A. Wander, N. Gura, H. Eberle, V. Gupta, and S. Chang, (2005)“Energy analysis for public-key cryptography for wireless sensor networks”, in the 3rd IEEE International Conference on Pervasive Computing and Communications, pp. 324–328, Italy.

### Author

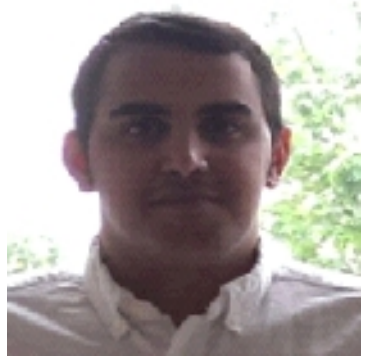

Saleh Almowuena received his B.Eng. in Computer Engineering from King Saud University, Saudi Arabia in 2006, and his M.A.Sc. in Electrical and Computer Engineering from the University of Victoria, Canada in 2010. Currently, he is a Ph.D. student in the School of Computing Science at Simon Fraser University, Canada. Saleh is working on improving the bandwidth usage and power consumption caused by Video-on-Demand services over LTE networks. His research interests also include the field of key management for Wireless Sensor Networks, the architecture of evolved-Multimedia Broadcast Multicast Service in LTE systems, and the behaviour model of users during multimedia streaming over mobile networks.